\documentclass[11pt, a4paper, twocolumn]{article}
\usepackage{inputenc}
\usepackage{hyperref}
\usepackage{amsmath}
\usepackage{physics}
\usepackage{amsfonts}
\usepackage{multirow}
\usepackage{color, soul}
\usepackage{orcidlink}
\usepackage{graphicx}
\usepackage{abstract}
\usepackage[affil-it]{authblk}
\usepackage[sorting=none, style=numeric-comp]{biblatex}
\bibliography{main}

\hypersetup{pdftitle={Long-lived Particles Anomaly Detection with Parametrized
Quantum Circuits}}

\title{Long-lived Particles Anomaly Detection with Parametrized
Quantum Circuits}

\author[1]{Simone Bordoni \orcidlink{0000-0002-9745-9189}}
\author[2]{Denis Stanev \orcidlink{0000-0002-7334-0159}}
\author[1]{Tommaso Santantonio}
\author[1]{Stefano Giagu \orcidlink{0000-0001-9192-3537}}
\affil[1]{La Sapienza University of Rome, Dep. of Physics, Rome, Italy}
\affil[2]{Gran Sasso Science Institute (GSSI), L'Aquila, Italy}

\begin{document}
\twocolumn[
\maketitle
\begin{@twocolumnfalse}
\begin{abstract}
We investigate the possibility to apply quantum machine learning techniques for data analysis, with particular regard to an interesting use-case in  high-energy physics.
We propose an anomaly detection algorithm based on a parametrized quantum circuit. This algorithm has been trained on a classical computer and tested with simulations as well as on real quantum hardware. Tests on NISQ devices have been performed with IBM quantum computers. For the execution on quantum hardware specific hardware driven adaptations have been devised and implemented.
The quantum anomaly detection algorithm is able to detect simple anomalies like different characters in handwritten digits as well as more complex structures like anomalous patterns in the particle detectors produced by the decay products of long-lived particles produced at a collider experiment.
For the  high-energy physics application, performance is estimated in simulation only, as the quantum circuit is not simple enough to be executed on the available quantum hardware. This work demonstrates that it is possible to perform anomaly detection with quantum algorithms, however, as amplitude encoding of classical data is required for the task, due to the noise level in the available quantum hardware, current implementation cannot outperform classic anomaly detection algorithms based on deep neural networks.\\ \vspace{10mm}
\end{abstract}
\end{@twocolumnfalse}]

\section{Introduction}
With the contemporary peak in interest regarding machine learning algorithms for their many applications in scientific research, we are also witnessing a rapid acceleration in the concurrent development of quantum computing. A combination of these two research fields has  led to the development of quantum machine learning  (QML) algorithms~\cite{qml, qml3, QONN, qml4}.
In this work,we propose a quantum version of a classic machine learning algorithm known as anomaly detection. This algorithm is implemented with an artificial neural network, in particular an autoencoder architecture~\cite{ae, ae2, ae3}. In quantum machine learning, the autoencoder is realised using parametrized quantum circuits~\cite{qml2, pqc, pqc2}.
First, we test a simpler version of the quantum algorithm in  an easier task involving a standard benchmark dataset in machine learning, the handwritten digits MNIST dataset. We then apply the technique to a more complex and interesting use-case, the identification of anomalous signatures inside a particle detector due to the decay of long-lived particles with macroscopic lifetimes.
The application of quantum machine learning to  high-energy physics is an interesting field that has been studied using QML simulators in some recent works~\cite{bjet, hep1, hep2, hep3, hep10, hep4, hep5, hep6, hep7, hep8, hep9}. In this paper, we present the first application of QML to the task of anomaly detection for long-lived particle identification, and also prove that the proposed variational quantum circuits  could be used on actual quantum hardware. The parametrized quantum circuits developed in this work are in fact simple enough to be tested on Noisy Intermediate Scale Quantum (NISQ) computers~\cite{nisq, nisq2, QuantumSupGoogle, nisq3}. The tests on real quantum hardware  have been implemented on IBM quantum computers~\cite{ibm}.
We stress here that the proposed algorithm is not meant, at this stage, to outperform the classical counterpart on classical data, but only to show that it is possible to use quantum variational algorithms for anomaly detection. 
QML algorithms will be employed in real applications only when less noisy qubits or effective error correction procedures will be available in digital quantum computers. The proposed algorithm may show future advantages for the analysis of quantum data~\cite{quantumdata}. In fact, it is really difficult to manage quantum data with classical hardware, due to the information size growing exponentially with respect to the number of qubits. On the other hand, a QML algorithm can directly analize quantum data, overcoming the problem of amplitude encoding (Sec.~\ref{sec_adapt}). \\

\noindent
The paper is structured in the following way: in Sec.~\ref{sec_background} the fundamental concepts underlying anomaly detection algorithms and parametrized quantum  circuits are briefly reviewed. Sec.~\ref{sec_main} describes the strategy used to develop the quantum algorithms and the performance estimated using  a simulator of the quantum circuits on classic hardware. In Sec.~\ref{sec_quantumhardware}, we describe the changes we have implemented in  the quantum circuit to be executed on the IBM\_hanoi\footnote{https://quantum-computing.ibm.com/services/\\resources?tab=systems\&skip=10\&system=ibm\_hanoi.} quantum computer, and discuss the results. Conclusions and future developments are presented in Sec.~\ref{sec_conclusion}.

\section{Background}\label{sec_background}
In this section we briefly recall the main elements needed to understand the proposed quantum anomaly detection algorithm.

\subsection{Anomaly detection algorithms} \label{sec_ada}

\begin{figure*}
\centering
\includegraphics[width=0.8\textwidth]{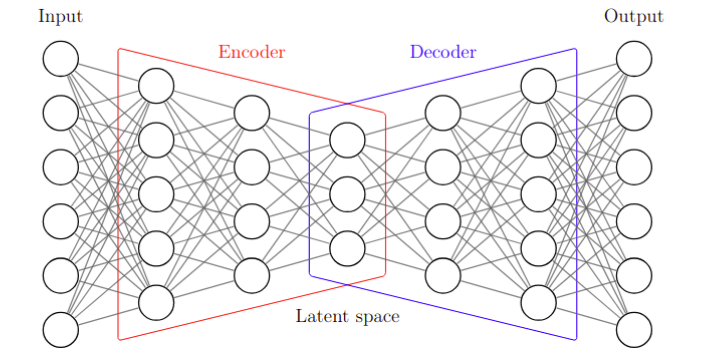}
\caption{Schematic representation of an autoencoder. The input data is compressed by the encoder, to a smaller number of features in the latent space. The decoder tries to reconstruct the original data from the compressed one.}
\label{fig:ae}
\end{figure*}

Anomaly detection describes a class of algorithms that aims at the identification of rare items, events or observations, which deviate significantly from the majority of the data and do not conform to a  well-defined notion of normal behavior. Anomaly detection has recently gained increasing interest in experiments at the Large Hadron Collider (LHC)~\cite{LHC}, as a viable machine learning approach to implement signal model-independent searches for new physics effects. The technique in fact requires only a precise prediction of the background (normal data) to train a classifier model to distinguish data from the background, without requiring a specific description of the new physics signal (anomalous data)~\cite{hep9}.\\
One way to implement an anomaly detection algorithm is to train a particular artificial neural network (ANN) architecture called autoencoder~\cite{ae, ae2, ae3},  utilized
in various applications of unsupervised learning. An autoencoder is composed of two main parts: an encoder and a decoder (Fig.~\ref{fig:ae}). The encoder compresses initial data down to a small dimension (called latent dimension). The decoder inverts the process to reconstruct the original data from the compressed representation. The parameters of the neural network are trained in order to  minimize the difference between the initial and reconstructed data. The loss function (also called reconstruction loss) is therefore a measure of how accurately the reconstructed data resembles the original. For anomaly detection, the autoencoder is trained only on data samples belonging to the normal event class (eg background). When the trained model is applied to new samples we expect the loss function to have different values for normal and anomalous data.
By choosing a threshold value for the loss function it is possible to classify an input based on whether its reconstruction loss lands above or below this threshold. The performance of the trained algorithm  is usually presented in terms of the Receiver operating characteristic (ROC) curve,  which shows the true positive rate as a function of the false positive rate at different classification thresholds~\cite{roc}, and in terms of the Area Under the ROC curve (AUC), which provides an aggregate measure of performance across all possible classification thresholds.

\subsection{Parametrized quantum circuits}\label{sec_pqc}

\begin{figure*}
\centering
\includegraphics[width=0.8\textwidth]{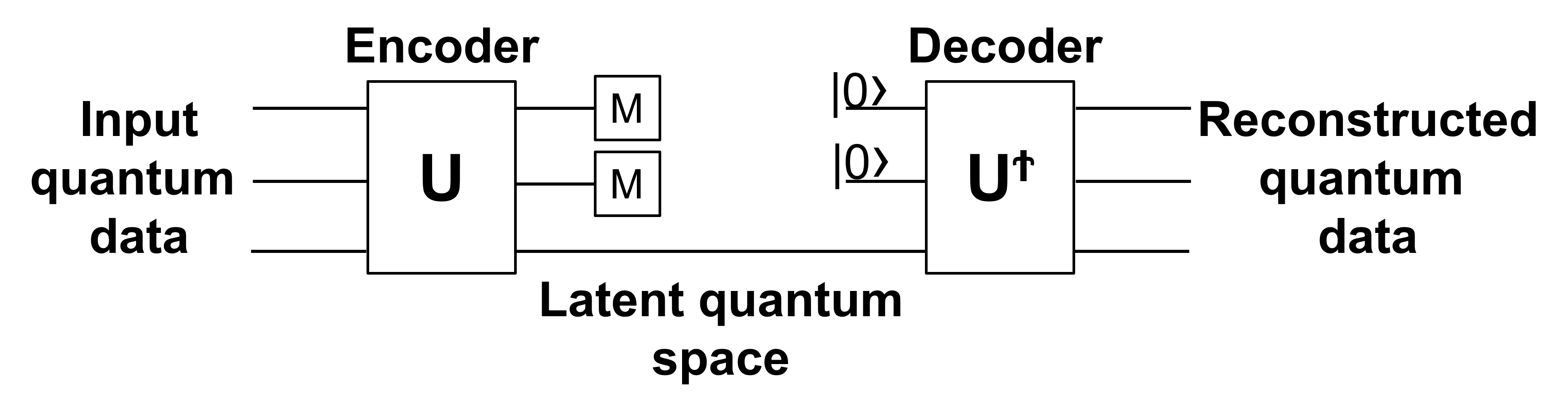}
\caption{Schematic representation of a quantum autoencoder. The encoder acts as a unitary transformation $U$ that tries to disentangle a certain number of qubits (force them to $\ket{0}$ state). In this way the initial quantum information is compressed into a latent quantum space. The decoder implements the inverse transformation $U^\dagger$ to reconstruct the original data. The loss function is taken as the measurement expectation value of the qubits that are eliminated after the compression.}
\label{fig:qae}
\end{figure*}

\begin{figure}
\includegraphics[width=0.5\textwidth]{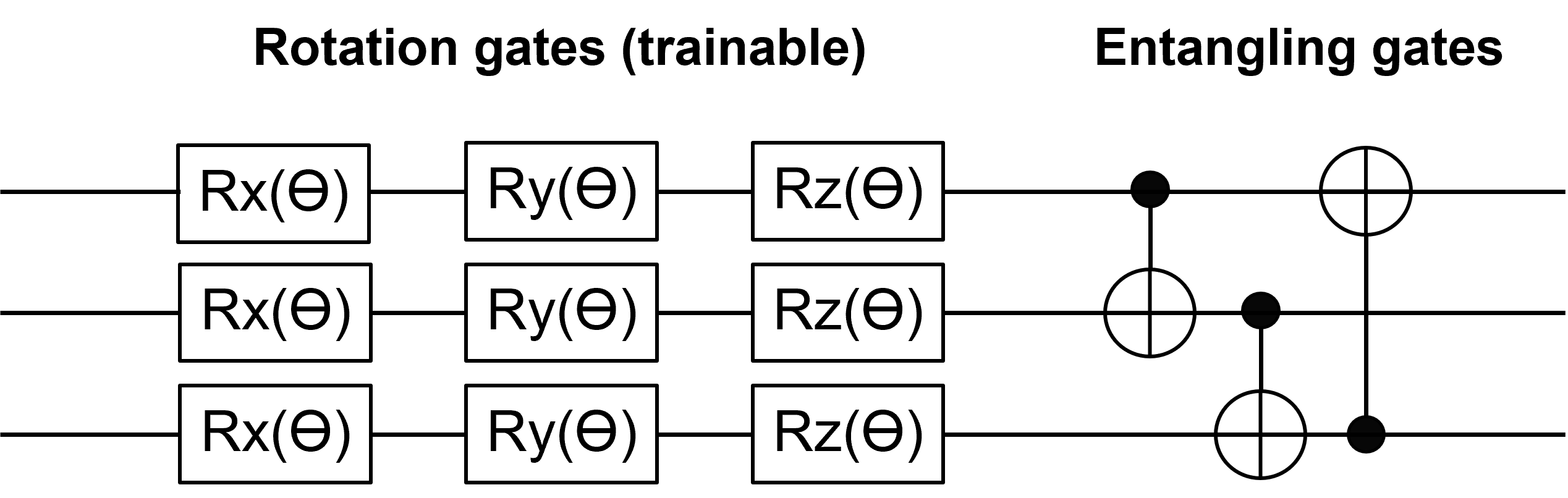}
\centering
\caption{Circuit representation for three qubits of one layer of the PQC used in this work. Single qubit rotation gates with trainable rotation angles are followed by an entangling layer made of C-NOT gates acting on neighbouring qubits.}
\label{fig:pqc}
\end{figure}

A parametrized quantum circuit (PQC) is a quantum algorithm that  depends on free parameters, and that can be used as the quantum counterpart of classical ANNs. In this kind of circuits, the input information is stored in the initial state of the qubits. It can be stored as the phase (phase encoding) or in the states amplitudes (amplitude encoding)~\cite{enc, enc2, enc3}. The initial state is transformed using rotation gates and entangling gates, usually controlled-not (C-NOT) gates~\cite{principles}. These gates can be organised in layers, in our circuit architecture one layer is composed of rotation gates ($R_x$, $R_y$, $R_z$) acting on all qubits followed by a series of C-NOT gates coupling  neighboring qubits (Fig.~\ref{fig:pqc}). The trainable weights are the angles of rotation gates and can be trained using the conventional stochastic gradient descent techniques via backpropagation adopted in the training of ANNs~\cite{back}.\\
A quantum circuit implements a unitary, thus invertible, transformation on the initial state. This represents a great advantage for the autoencoder architecture, as the decoder can be taken as simply the inverse of the encoder quantum circuit (Fig.~\ref{fig:qae}). In order to compress information the encoder circuit has to disentangle and set to zero state a given number of qubits~\cite{qae}. The loss function is thus taken as the expected measurement values of these qubits. In this way, for the training of the circuit, it is necessary only the encoder.
For the simulation and training of the PQC we have used the QIBO~\cite{qibo} library that can be easily integrated with Tensorflow~\cite{tens} for automatic differentiation.

\section{Simulation on classic hardware}\label{sec_main}
In order to find the best parameters of the proposed anomaly detection algorithm the first tests have been carried out with simulations on classical hardware.  The anomaly detection task can be solved, with satisfactory discriminative power, by requiring a PQC with enough expressive power.
This can be achieved by increasing the depth of the circuit. However, increasing the depth of the circuit may lead to difficulties in finding a suitable minimum in the loss function during the training, mainly due to the presence of barren plateaus~\cite{bp,bp2}. To reach an acceptable trade-off between the two effects, a detailed study of the different possible quantum gate topologies that could be used in the circuit was carried out. The algorithm has been applied at first for the recognition of anomalous handwritten digits (Sec.~\ref{sec_mnist}), and then for the  high-energy
physics problem (Sec.~\ref{sec_hep}). In both cases, the quantum encoding of the classical input data has been implemented using amplitude encoding. In this way, it is possible to encode a number of features that grows exponentially with the number of qubits in the quantum circuit. A drawback with the amplitude encoding, on the other hand, is in the state preparation that requires an exponential number of gates~\cite{sp} with respect to the number of qubits. State preparation is a current and still open problem on NISQ devices.

\subsection{Simple use-case: handwritten digits}\label{sec_mnist}

\begin{figure}
\includegraphics[width=0.25\textwidth]{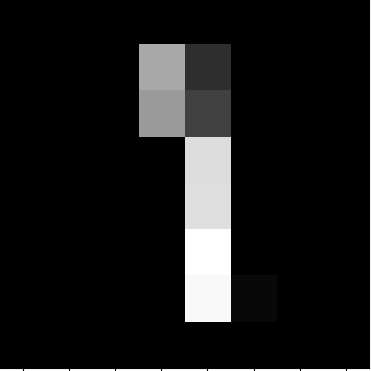} \hspace{1cm}
\includegraphics[width=0.25\textwidth]{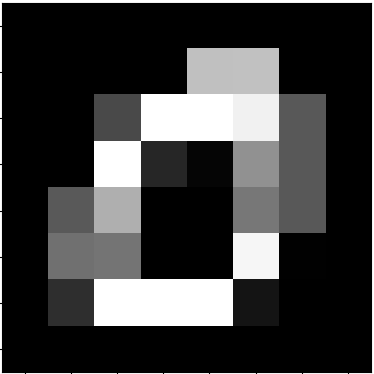}
\centering
\caption{Example of images for the  "zero" and "one"
digits for the MNIST handwritten digits dataset. Images are compressed to 8$\times$8 pixels.}
\label{fig:fig_mnist}
\end{figure}

Anomaly detection on handwritten digits has been carried out on the MNIST dataset. We define handwritten  "zero"
digits as normal data, and  "one" digits as anomalous data. Examples of two images from the MNIST dataset are shown in Fig.~\ref{fig:fig_mnist}. 
The original MNIST images are compressed down to 8$\times$ 8 pixels, every pixel is an integer number from 0 to 255 (8 bit grayscale).
In order to encode the classical data, the images are flattened to obtain feature vectors composed of 64 elements.
These vectors are then normalised so that they can be encoded in the state amplitudes of 6 qubits using amplitude encoding (Sec.~\ref{sec_pqc}).
We tested the PQC with a different number of layers and number of compressed qubits.
\begin{figure*}
\centering
\includegraphics[width=0.9\textwidth]{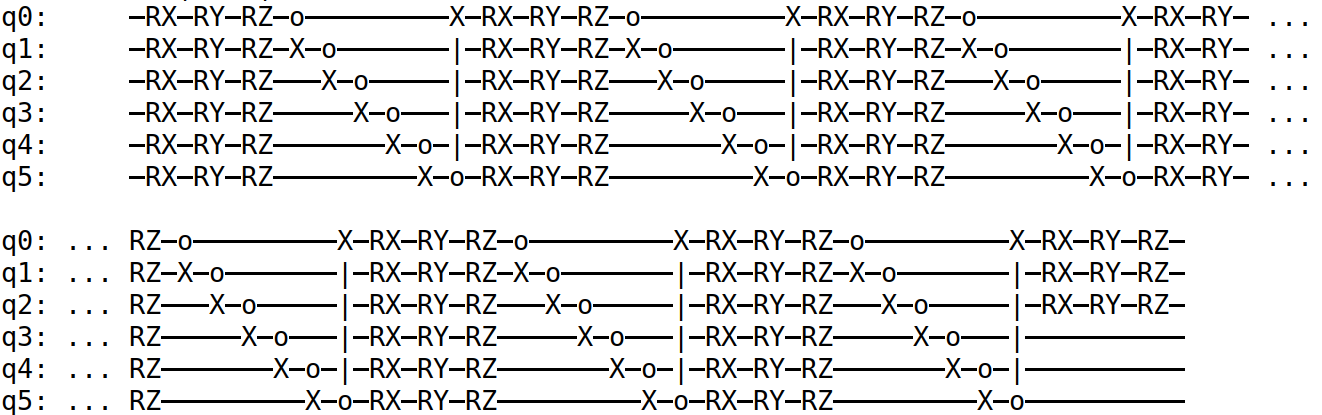}
\hspace{2mm}
\caption{Circuit representation of the quantum encoder used for anomaly detection of handwritten digits. The encoder circuit is composed of six qubits and six layers. At the end of the quantum circuit the first three qubits are measured in order to compress information on the three remaining qubits.  The loss function is computed as the sum of the probabilities of having any of the measured qubits in the ground state.}
\label{fig:circ_mnist}
\end{figure*}
\begin{figure*}
\centering
\includegraphics[width=0.8\textwidth]{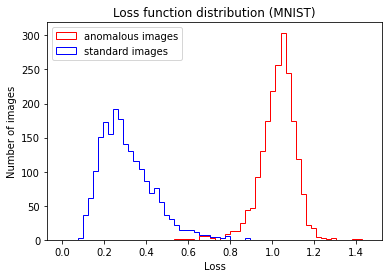}
\caption{Quantum autoencoder loss function values distribution. The graph has been made using $2000$ normal data images (zeros) and $2000$ anomalous data images (ones) from the MNIST dataset.}
\label{fig:dist_mnist}
\end{figure*}
The number of layers has been varied from 4 to 8. This interval has been chosen because circuits with less than 4 layers don't have enough expressive power to solve the task. On the other hand, incrementing the number of layers to more than 8 makes the training procedure complicated because of barren plateaus.
The number of compressed qubits has been varied from 1 to 4. In this range, the decoder is able to reconstruct the original images with a good precision.
The performances of the different circuit architectures have been compared using the AUC value (Sec.~\ref{sec_ada}). The best configuration has been found with six layers and three compressed qubits.  This procedure has been repeated in order to find the best entangling gates ansatz. We have tested different C-NOT configurations, the one that produced the best performance is reported in Fig.~\ref{fig:circ_mnist}. In order to improve the performance, rotation gates with trainable parameters were added at the end of the encoder circuit for the three compressed qubits. It is worth  noticing here that the chosen configuration requires only nearest neighbour connectivity for six qubits placed in a ring topology. We also verified that is possible to preserve  a good performance  by reducing the number of layers from six to four. We leverage these two properties during the implementation on quantum hardware of the PQC  (Sec.~\ref{sec_adapt}), to  minimize the disruption of the algorithm performance due to the noisy device. 
For the training of the circuit, a dataset of $5000$ images of zero handwritten digits has been employed. 
As the first three qubits are forced to $\ket{1}$ state in order to apply the data compression, the loss function is the sum of the probabilities of having any of these three qubits in the ground state.
Training has been performed for 20 epochs using Adam optimizer ~\cite{adam}, with a dynamic learning rate that spans from 0.4 in the first epochs to 0.001 in the last ones. This variable learning rate has helped reducing the problem of barren plateaus.
The number of epochs is sufficient to reach a plateau in the loss function. No overfitting has been observed during the training process, thus no early stopping has been required. For the training we have used batches composed of 20 samples (250 steps per epoch).\\ To test the anomaly detection algorithm after the training phase, we have used 2000 normal data images not used in the training and 2000 anomalous data images. Figure~\ref{fig:dist_mnist} shows the loss distribution for the two test datasets.
It is possible to observe a clear separation, with an average loss value for normal data of 0.307 and 1.026 for anomalous data, with an average Root Mean Squares (RMS) of the two distributions of 0.110 and 0.066 respectively.

\subsection{High-energy physics use case}\label{sec_hep}

\begin{figure*}
\includegraphics[width=\textwidth]{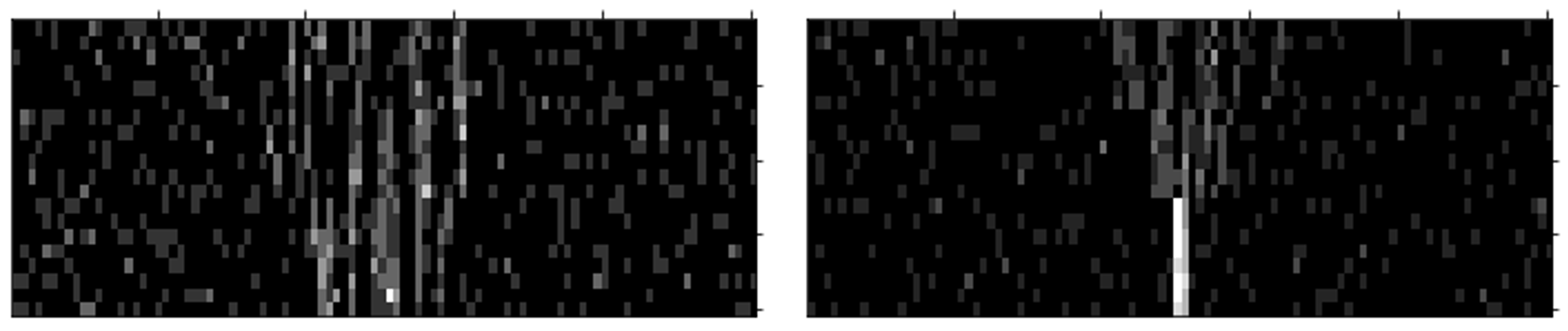}
\centering
\caption{A normal (left) and an anomalous (right) event used for quantum anomaly detection applied to long-lived particles detection. Normal and anomalous data are respectively prompt and displaced decays in multi-muons (from two to ten muons). Hits produced by the muons are reconstructed in a toy simulation of the ATLAS MDT chambers, that includes random hit background noise mimicking expected ATLAS phase-2 noise. Data are represented in the form of 
100$\times$20 pixels  images.}
\label{fig:samples}
\end{figure*}

An interesting use-case for an high-energy physics application of an anomaly detection algorithm is the identification of anomalous patterns in the triggers system of a collider experiment.  As a benchmark scenario for this work, we investigated the specific case of long-lived exotic particles predicted in new physics models with a secluded sector, like in~\cite{DARKPHOTONS, HVALLEY}. New particles predicted by theories beyond the Standard Model can generically have lifetimes that are long compared to Standard Model particles at the weak scale. When produced at the LHC, such long-lived particles can decay far from the primary interaction vertex and possibly themselves interact with the detector material, leading to a plethora of possible detector signatures. Such signatures are distinctly different from those associated with traditional searches for prompt particles, and requires dedicated reconstruction and identification algorithms, especially in the trigger systems of the collider detectors.  As an example case, it is possible to think about the high-level muon trigger system of the ATLAS experiment~\cite{atlas}, where the computational demands and the lack of flexibility of traditional secondary vertex reconstruction algorithms in the muon spectrometer~\cite{PERF-2013-01}, restrict their applicability only to the offline reconstruction~\cite{PhysRevD.101.052013}. This prevents their use in the trigger system of the experiment, limiting the discovery potential of the experiment for this kind of new physics searches.
The ATLAS experiment muon system aims to collect all the particle hit information from different sub-detectors to find muon candidates in a given sector (i.e. a solid angle region of the detector). In this proof of concept study we restrict our attention only to the barrel region of the muon spectrometer and to the Muon Drift Tube precision detector (MDT). The trigger algorithm tries to find patterns of hits consistent with the presence of muons originating from the same production vertex. One can think to arrange the MDT detector hits into image-like objects, to be used as input for ML algorithms particularly suitable to find patterns like the muon tracks in this test scenario. Using the published ATLAS detector geometry and resolution as well as its magnetic field map~\cite{atlas}, it is possible to generate toy events with muon tracks, from the decays of a non-interacting neutral long-lived particle at different decay lengths from the primary proton-proton interaction vertex. 
The images are produced initially with a pixel resolution that roughly corresponds to the MDT detector segmentation, each horizontal pixel corresponding to the center of a MDT tube, and each vertical pixel corresponding to one of the layer of MDT tube in the sector. The simplified simulation is only parametric, associates a binary value to each pixel (zero or one) depending if one of the muons from the long-lived particle decay went through the tube, and then applies position smearing based on the published ATLAS detector performance.
Random hit background emulating the typical noise rate conditions during the LHC runs has been also included in the simulation. The background noise is generated only accounting for the average hit rate in the spectrometer MDTs, therefore it does not consider correlated backgrounds. We did not aim here to perfectly reproduce the experimental conditions, but to give a proof of principle of the anomaly detection algorithm in the context of a high-energy physics experiment.
The images obtained in the simplified simulation are too large (20$\times$333 pixels) to be encoded as quantum input in the available quantum computers, therefore they have been compressed with a pooling operation to reduce their size to 20$\times$100 pixels. The pixels of the compressed image contain the integral of the original pixel content pooled and therefore the image appears as a grayscale and no longer as a binary image.
Events are generated with a single particle gun generator producing the decay products of an hypothetical dark particle promptly decaying in multi muons in the primary vertex of the detector, and then translated at different decay lengths. The momentum distribution of the particle mimic the typical expected one for a dark-photon in the Falkowski–Ruderman–Volansky–Zupan model ~\cite{DARKPHOTONS}, the mass of the long-lived particle has been randomly chosen in the range [0.5, 5] GeV.
Two datasets are generated, one corresponding to prompt to short decay length decays in multi-muons (from two to ten muons), with radial decay lengths uniformly distributed between 0.0 and 20.0 cm from the primary interaction vertex, and one corresponding to very displaced decays in multi-muons, with radial decay lengths uniformly distributed between 250.0 and 450.0 cm.
Data are conveniently represented in the form of images of dimension 100$\times$20 pixels (Fig.~\ref{fig:samples}). 
The encoding procedure used for this dataset is the same as the one described for the MNIST dataset (Sec.~\ref{sec_mnist}). In this case, 11 qubits are required for amplitude encoding.
In order to find the best PQC ansatz, the same procedure described in Sec.~\ref{sec_mnist}  has been repeated.
For this task we have tested PQCs with a number of layers varying from 6 to 10 and a number of compressed qubits varying from 1 to 4.  Choosing a number of layers between this interval represents a good trade-off between the expressive power and the problem of barren plateaus.
The best performance (AUC value) has been obtained with 8 layers and 3 qubits compression.
Training has been performed on a dataset of 8000 normal data images for 60 epochs.  The number of epochs is sufficient to reach a plateau in the loss function. Also in this case, no overfitting has been observed during the training process, thus no early stopping has been required. For the training we have used batches composed of 20 samples (400 steps per epoch). The optimizer and learning rate are the same used in Sec.~\ref{sec_mnist}. In order to mitigate barren plateaus we have also tried to train the circuit layer by layer  as suggested in~\cite{layerwise} but the final result was worse than a single all-layers training.\\

\begin{figure*}
\hspace{-12mm}
\includegraphics[width=0.56\textwidth]{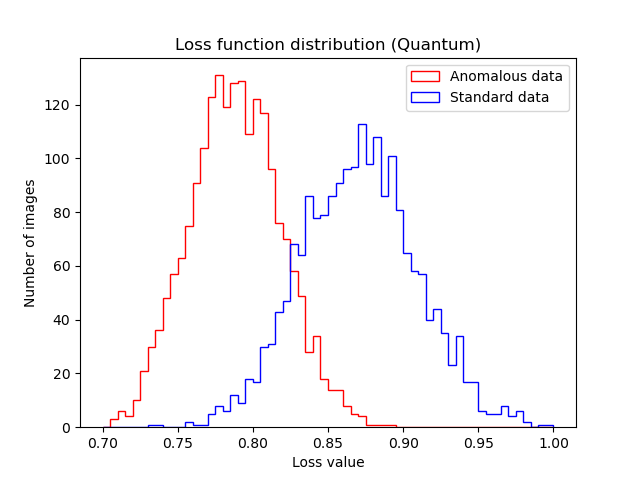}
\hspace{-11mm}
\includegraphics[width=0.56\textwidth]{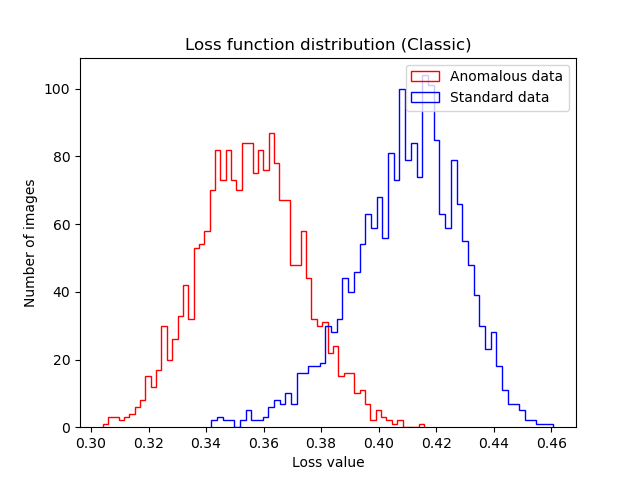}
\hspace{-17mm}
\centering
\caption{Loss function values distribution for the quantum anomaly detection algorithm (left) and the classic counterpart (right). The graphs have been made using 2000 normal data images (decays that happen just as the particle enters the detector) and 2000 anomalous data images (decays that happen after the particle has travelled inside the detector).}
\label{fig:dist_hep}
\end{figure*}

\begin{figure*}
\includegraphics[width=0.75\textwidth]{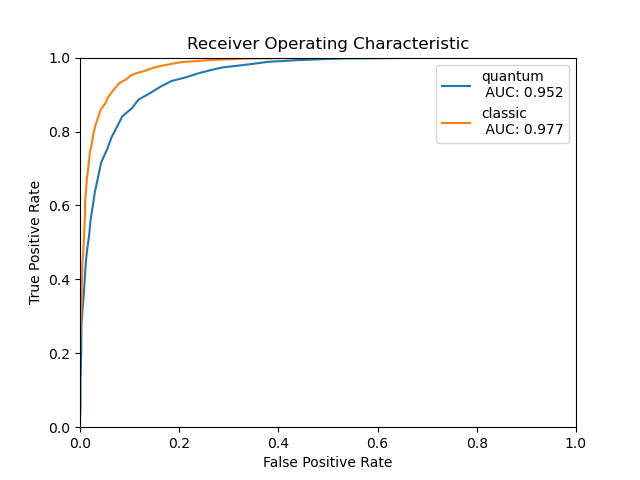}
\centering
\caption{ROC curve and AUC for quantum anomaly detection algorithm (blue) and the classic counterpart (orange).}
\label{fig:roc_hep}
\end{figure*}

\noindent
In order to benchmark the performance of the quantum algorithm we have made a comparison with a classic anomaly detection algorithm. The standard autoencoder has been implemented with a convolutional neural network (CNN). The Encoder CNN is composed of three convolution layers followed by a dense layer and compresses the input features down to a latent space of dimension 5.
The decoder CNN is composed of a dense layer followed by three transpose convolution layers that restore the original input dimension. The total number of parameters of this neural network is about $7.9\times10^3$. The loss function is the binary cross entropy between the input and the reconstructed image. Training has been performed on the same dataset used in the quantum case, for 60 epochs and using Adam optimizer.
The final performance evaluation has been carried out on 2000 normal data images and 2000 images of anomalous decays for both the quantum and classic  algorithms.
The loss function distributions for the normal and anomalous datasets are reported in Fig.~\ref{fig:dist_hep}. It is possible to observe a separation in the loss function distribution for anomalous and normal data images in both cases. For the quantum algorithm, the resulting average loss for normal events is $0.870$ while  the average loss for anomalous events is $0.788$. The RMS of the two distributions are respectively $0.038$ and $0.030$. For the classic algorithm, the resulting average loss for normal events is $0.409$ while the average loss for anomalous events is $0.355$. The RMS of both distributions is $0.017$. Note that as the loss functions are different for the classic and quantum case only the relative separations can be compared. 
It is worth  noticing here that, for both algorithms, the anomalous data show a lower average loss than normal data. This is due to the fact that anomalous images have a simpler structure with hits distributed in a narrow cone with respect to the normal images, and thus are easier to compress.
For a better benchmark, Fig.~\ref{fig:roc_hep} shows the ROC curve and AUC comparisons between the classic autoencoder and the quantum model.  The quantum algorithm does not reach the same level of performance as the classical equivalent. This is due to the reduced expressive power of the implemented quantum circuit, imposed by the constraints on the number of usable qubits and gates for the simulation. In any case, the quantum algorithm is already very close to the classical one, and is expected to significantly improve, on quantum hardware implementation, with the availability of the new generation of low noise quantum circuits which has been announced to be available soon.~\footnote{See for example the IBM Roadmap to quantum advantage: https://www.ibm.com/quantum/roadmap.}


\section{Test on quantum hardware} \label{sec_quantumhardware}
The execution of quantum circuits on NISQ devices, as stated in Sec.~\ref{sec_main}, is difficult even on  state-of-the-art quantum devices. The main problems come from the high error rate of quantum gates, especially C-NOT gates that are fundamental to generate entanglement between qubits. Another important limitation comes from the connectivity in the architecture of quantum computers. In fact, it is not possible to apply C-NOT gates between all possible pairs of qubits. If an interaction between not connected qubits is required, it is necessary to use SWAP gates to invert the quantum states of two qubits. As each SWAP gate is composed of three C-NOT gates~\cite{principles},  the use of noisy gates is further increased. State-of-the-art quantum computers offer connectivity only between neighbouring qubits, arranged in linear or circular structures. The architecture of the quantum computer used in this work (IBM\_hanoi) is reported in Fig.~\ref{fig:hanoi}.

\subsection{Adaptation to quantum hardware}\label{sec_adapt}

\begin{figure}
\includegraphics[width=0.5\textwidth]{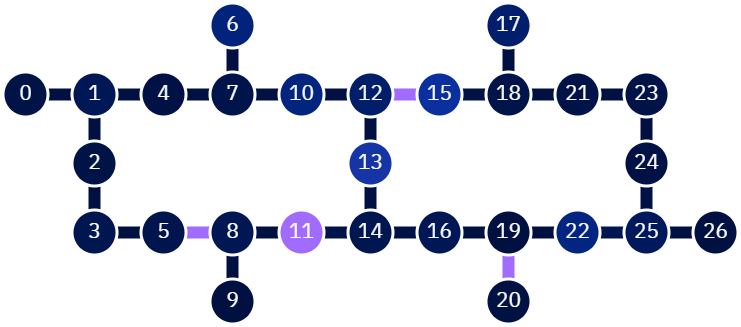}
\centering
\caption{Architecture of IBM\_hanoi quantum computer. The C-NOT connectivity is reported. The colours of single qubits and their connections represent respectively the single qubit readout assignment error and C-NOT error probabilities. Darker colours represent a lower error probability, in a range between 5.9$\times$10$^{-3}$ and 9.8$\times$10$^{-2}$ for readout error and 3.3$\times$10$^{-3}$ and 1 for C-NOT gates. Data from calibration on $19/10/2022$.}
\label{fig:hanoi}
\end{figure}

\begin{figure*}
\centering
\includegraphics[width=0.8\textwidth]{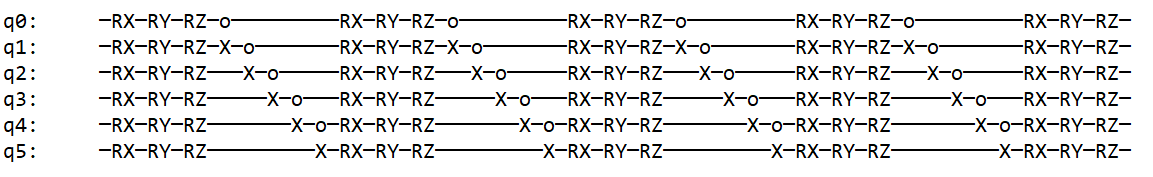}
\caption{Parametrized quantum circuit used for approximated amplitude encoding. The circuit is composed of four layers made of rotation gates and five C-NOT gates plus a final layer composed only of rotation gates.}
\label{fig:ampencoding}
\end{figure*}

With these limitations is impossible to achieve any significant performance on the IBM quantum hardware for the anomaly detection circuit we have developed.  In order to make the algorithm work, some changes and a careful adaptation had to be implemented to reduce the complexity and size of the PQC. Given the consequent reduction in the expressive power of the model, we decided to focus only on the simplest use-case of the handwritten digits (see Sec.~\ref{sec_mnist}) for the quantum hardware test.
We describe here the changes to the quantum circuit in order to solve the two problems discussed in the previous Sec., the amplitude encoding and the C-NOT connectivity.\\
For the connectivity problem it is important to notice that the PQC proposed in Sec.~\ref{sec_pqc} requires only neighbouring qubits interaction if the qubits are arranged in a circular topology. However, in our quantum hardware (Fig.~\ref{fig:hanoi}) it is not possible to find six qubits arranged in a ring. By removing the last C-NOT gate on each entangling layer only neighbouring qubits interactions are required for qubits arranged on a line. This allows to remove all SWAP gates from the circuit, thus reducing significantly the total number of C-NOT gates.\\
Moreover we decided to use only four layers for the encoder circuit.
With these changes, the expressive power of the PQC is reduced but still sufficient to detect anomalies. On noisy simulations the performance of this PQC outperformed the one with 6 layers employed in Sec.~\ref{sec_mnist}.\\

\noindent
Amplitude encoding is a state preparation procedure that allows to transform the standard initial state (all qubits in the ground state) into a state that encodes the input data for the quantum algorithm. This procedure is necessary only to analyse classical data with quantum hardware. Implementing amplitude encoding on quantum hardware requires a number of C-NOT gates that grows exponentially in the number of qubits (Sec.~\ref{sec_main}). For a circuit of 6 qubits more than one hundred C-NOT gates are required, this adds too much noise to the final result. To overcome the problem we developed another PQC designed to provide a good approximation of the exact amplitude encoding while using a reduced number of gates. This circuit is trained to transform the initial ground state into a state that approximates amplitude encoding. The parameters of the circuit have been chosen by minimising the mean squared error between the output state of the circuit and the target state. For this procedure one circuit  has to be trained for each normal or anomalous data that we want to encode.
The PQC ansatz is composed of four layers with the C-NOT gate disposition described before for the encoder (Fig.~\ref{fig:ampencoding}). Moreover, a final layer composed only of rotation gates has been added to increase expressive power and improve performance.
Each training has been carried out for 15 epochs, with each epoch composed of 100 training steps, using Adam optimizer and a learning rate of 0.01. In order to avoid barren plateaus we have started each training using the parameters of the previous PQC.
Looking at the final loss values, we found out that the encoding of anomalous data (1 digits) is easier than the encoding of normal data (0 digits).
In order to avoid possible bias introduced by bad amplitude encoding, we selected only the data encoded with a loss smaller than 0.1. We have trained approximated amplitude encoding with these characteristics for about 200 normal and 200 anomalous data.

\subsection{Results}

\begin{figure}
\includegraphics[width=0.5\textwidth]{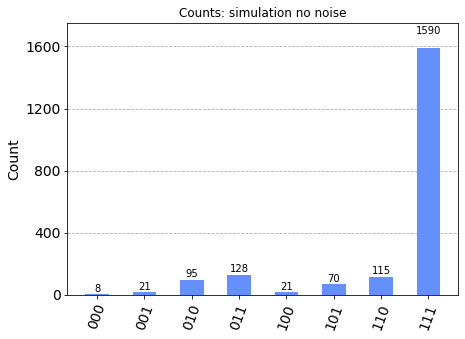}
\includegraphics[width=0.5\textwidth]{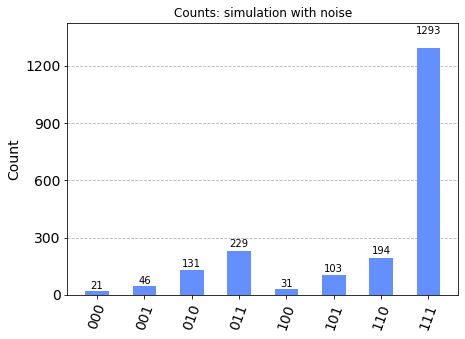}
\includegraphics[width=0.5\textwidth]{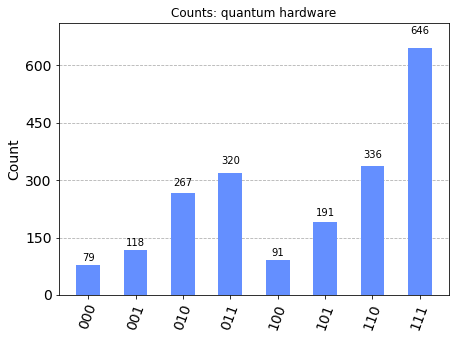}
\centering
\caption{Counts distribution of 2048 shots for a simulated circuit with no noise (top) a simulated circuit with realistic noise (center) and a noisy quantum circuit (bottom).}
\label{fig:shots}
\end{figure}

\begin{figure*}
\centering
\hspace{-5mm}
\includegraphics[width=0.50\textwidth]{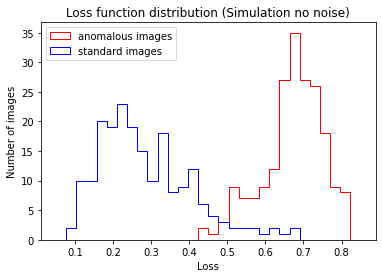}
\includegraphics[width=0.50\textwidth]{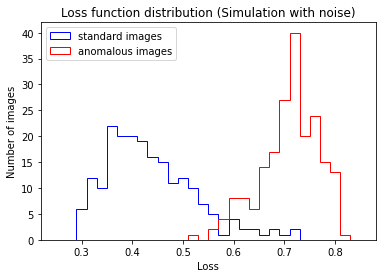}
\hspace{-5mm}
\includegraphics[width=0.50\textwidth]{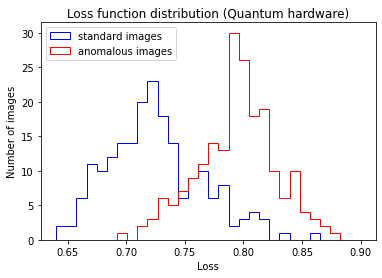}
\caption{Quantum autoencoder loss function values distribution. Simulated circuits with no noise (top left), simulated circuits with noise (top right) and a noisy quantum circuits (bottom). The graph has been made using 200 normal data and 200 anomalous data, with 2048 shots each circuit. The loss function is one minus the probability of the $\ket{111}$ state.}
\label{fig:lasttest}
\end{figure*}

\begin{figure*}
\includegraphics[width=0.7\textwidth]{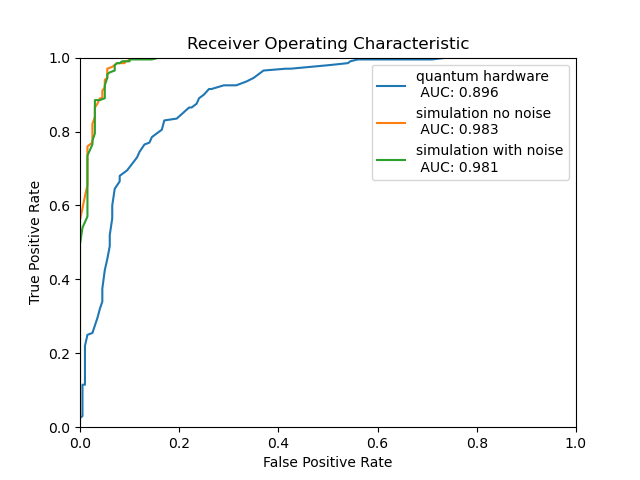}
\centering
\caption{ROC curves and AUC for anomaly detection, simulated circuits with no noise (orange), simulated circuits with noise (green) and noisy quantum circuits (blue).}
\label{fig:lastroc}
\end{figure*}

The final test on quantum hardware has been carried out using a circuit composed of two parts, the approximated amplitude encoding circuit and the encoder (Sec.~\ref{sec_adapt}). The parameters of the encoder have been trained with simulations on classic hardware in the same way as described in Sec.~\ref{sec_mnist}.\\
The parameters of the approximated amplitude encoding circuit have been trained on each dataset as described in the previous section. 200 samples of normal data and 200 samples of anomalous data have been tested on the quantum hardware. Each circuit has been executed with 2048 shots.\\
Figure~\ref{fig:shots} reports a comparison of the measurement counts of the circuit (only the first three qubits are measured) for a normal sample. On the left plot, the simulated circuit without noise is shown, on the center plot the simulated circuit with noise. For the noise model we have used the calibration values of IBM\_hanoi taken on $19/10/2022$ \footnote{ The calibration can be accessed through IBM’s qiskit library: https://qiskit.org.}. The plot on the right reports the counts for the circuit executed on the quantum hardware. As explained in Sec.~\ref{sec_mnist}, the loss function,  which the circuit has to minimize, measures the sum of the probabilities of the three compressed qubits to be in the ground state. Thus, if the algorithm is working correctly, we expect the measured qubits to be mainly in the $\ket{1}$ state. The counts distribution is peaked, as expected, on the $\ket{111}$ state in all cases. The real quantum circuit, however, clearly shows a higher level of noise shown by higher counts for states different from $\ket{111}$. Moreover, the noise reported by IBM's simulation is lower than the one on real quantum hardware.  This is probably due to
the difficulty in reproducing a realistic noise model. In a real quantum circuit there are many sources of noise, besides the readout measurement error and the gate error, so it is complicated to keep track of them all. Moreover, the noise parameters, obtained from calibration, change over time. This makes them no longer reliable if the circuit is not executed right after the calibration, like in this case.
The noise makes the loss function distributions for normal and anomalous data almost indistinguishable. However, by observing the counts, we noticed that it is possible to improve the performance of anomaly detection by using as loss function one minus the probability of the $\ket{111}$ state.  This loss function is different from the one previously employed. In fact, it can be easily observed that: $1-p(\ket{111})\neq p(\ket{0}_1)+p(\ket{0}_2)+p(\ket{0}_3))$. This new loss function gives better results in the presence of noise because the state $\ket{111}$ is the one with the highest output probability. Output state probabilities are computed as the number of counts, for that specific state, divided by the total number of counts. Thus, probabilities of states with higher counts are less affected by the noise, compared to the probabilities of states with a lower number of counts.\\
The distributions for normal and anomalous data for this loss function are reported in Fig.~\ref{fig:lasttest} for simulated circuits with no noise (top left), simulated circuit with noise (top right) and real quantum circuit (bottom). For a better comparison Fig.~\ref{fig:lastroc} reports the ROC curves and the AUC for the three cases. It is possible to observe a significant separation between normal and anomalous data in the presence of noise, although with a clear degradation in the case of the execution on real quantum hardware (hardware vs simulation AUC: 0.896 vs 0.983).\\
It is interesting to notice that in the case of simulation, the ROC curves for simulated circuits with or without noise almost overlap even if the loss values for these two cases have different distributions. A low level of noise just shifts the losses to higher values but does not reduce the discriminative performance of the algorithm.

\section{Conclusions}\label{sec_conclusion}
Quantum machine learning is a newborn topic with many possible algorithms still to be explored. In this work we proved that it is possible to do anomaly detection for long-lived particles searches in a high-energy physics experiment using parametrized quantum circuits. Theoretically, without considering noise limitations, PQCs are powerful enough to distinguish anomalies in complex patterns like the ones obtained in a muon spectrometer of a hadron collider experiment. With the nowadays available NISQ devices it is only possible to execute very simple circuits on quantum hardware. These circuits must be adapted to the hardware limitations and can be used only for simpler tasks like anomaly detection of handwritten digits. However, as quantum computers are improving rapidly~\cite{road, road2}, we expect, in the near future, to use PQCs also to solve complex tasks in particle physics. At that time will be possible to verify if, with  fault-tolerant
quantum computation, quantum machine learning will outperform classical machine learning algorithms on the analysis of classical data. 
Another possible direction to explore is the possibility to feed the quantum algorithms directly with quantum input data, thus avoiding the quantum data encoding step. Since the research and development of quantum sensing in particle detectors is a rapidly developing sector~\cite{qt1,qt2,qihep}, this could lead to identify a possible advantage from the use of quantum machine learning algorithms, of the type proposed in this work, in a not too distant future.\\

\noindent
\textbf{Acknowledgments}\\
This work has been partially supported by ICSC - Centro Nazionale di Ricerca in High Performance Computing, Big Data and Quantum Computing, founded by European Union - NextGeneration EU.

\newpage
\onecolumn
\printbibliography

\newpage
\section*{Abbreviations}
The following abbreviations are used in this manuscript:\\

\noindent 
\begin{tabular}{@{}ll}
QML & Quantum machine learning\\
NISQ & Noisy intermediate scale quantum\\
ANN & Artificial neural network\\
PQC & Parametrized quantum circuit\\
ROC & Receiver operating characteristics\\
CNN & Convolutional neural network\\
AUC & Area Under the roc curve\\
LHC & Large hadron collider\\
MDT & Muon drift chamber\\
RMS & Root mean square\\
ML & Machine learning
\end{tabular}

\end{document}